\documentclass[twocolumn,a4paper]{article}
\usepackage[utf8]{inputenc}
\usepackage{xcolor}
\usepackage[colorinlistoftodos]{todonotes}
\usepackage[caption=false]{subfig}
\usepackage{multirow}
\usepackage{import}
\usepackage{authblk}
\usepackage[pdftex,pagebackref,plainpages=false,linktocpage]{hyperref}
\def \TS {Wiselib TupleStore}
\def \SPO {\textit{(Subject Predicate Object)} }
\def \SPITFIRE {SPITFIRE }

\begin{document}

\title{The \TS: A Modular RDF Database for the Internet of Things}
%\title{The \TS{}: A Modular RDF Database for the IoT}

\author{Henning Hasemann, Alexander Kröller}
\affil{
	TU Braunschweig,
	Germany, \texttt{\{h.hasemann,a.kroeller\}@tu-bs.de}
}
	
	%21 Lower Kent Ridge Rd\\
\author{Max Pagel}
\affil{
	National University of Singapore,
	Singapore,
	\texttt{max.pagel@nus.edu.sg}
}

\maketitle

\begin{abstract} The Internet of Things movement provides
self-configuring and universally interoperable devices. While such devices are
often built with a specific application in mind, they often turn out to be
useful in other contexts as well. We claim that by describing the
devices' knowledge in a
universal way, IoT devices can become first-class citizens in the Internet.
They can then exchange data between heterogeneous hardware, different
applications and large data sources on the Web.  Our
key idea --- in contrast to most existing approaches --- is to not
restrict the domain of knowledge that can be expressed on the device in any
way and, at the same time, allow this knowledge to be machine-understandable and
linkable across different locations.
We propose an architecture that allows to connect embedded devices to the
Semantic Web by expressing their knowledge in the Resource Description
Framework (RDF).  We present the \textit{Wiselib TupleStore}, a modular
embedded database tailored specifically for the storage of RDF. The Wiselib TupleStore is
portable to many platforms including Contiki and TinyOS and allows a variety
of trade-offs, making it able to scale to a large variety of hardware scenarios.  We
discuss the applicability of RDF to heterogeneous resource-constrained devices
and compare our system to the existing embedded tuple stores Antelope and
TeenyLIME.  \end{abstract}

%\IEEEpeerreviewmaketitle

% Feedback from Matteo Ceriotti:
% - TupleStore not enough for a general storage mechanism (TennyLime & co
%   better)
% -> Show its use for RDF maybe later illustrate fleixbility
% - network view interesting
% - provide complete view of the system

\section{Introduction}\label{sec:intro}

We are witnessing the evolution of Wireless Sensor Networks (WSN) into
the Internet of Things (IoT), bridging the world of
resource-constrained embedded devices to powerful machines and vast
data clouds in the internet. Analysts predict enormous numbers of
devices being connected to the Internet, for example,
ABIresearch\footnote{https://www.abiresearch.com/press/more-than-30-billion-devices-will-wirelessly-conne}
estimates 30 billion devices by 2020.  To make this growth possible,
hard- and software for networked embedded devices must become an
easy-to-use commodity, especially when it comes to building complex
applications.

%   This fusion brings a number of challenges and
% opportunities with it: Embedded devices are heterogeneous in both hard- and
% software and often designed with specific applications in mind. A sound
% integration with the internet however demands universal ways of exchanging
% data not only between the devices and more powerful machines on the current
% internet but also between the diverse and heterogeneous embedded devices.  In
% parallel to this consolidation of networks, we observe a rapidly growing
% number of devices and deployments, especially targeting end-users.  This
% yields the demand for a class of devices that is installable in a plug \& play
% manner, easily configurable, connects to the Web and can be used for unforeseen
% future applications.  At the same time this opens up the opportunity of a
% totally new class of applications that combine data from the various sources
% found in the Web with live sensor data from embedded IoT devices and thus has
% an unpreceeded pool of knowledge at its disposal.

To reach this goal, several demands have to be met: On the
lowest layer, embedded devices have to be able to exchange messages
and to identify each other using an agreed set of protocols, including
a connection with the existing Web and its data pool.  This is conducted---beyond other approaches---by the
elaboration of standardized protocols like
6LoWPAN~\cite{Mulligan:2007:ARC:1278972.1278992} and
CoAP~\cite{citeulike:9541809}, that map in a straightforward way to IPv6 and
RESTful Web Services. %; we thus focus on the data layer integraton

However, this is hardly enough: While these protocols define the means
of communication, that is, \textit{how} data is transferred, they
leave it open to the application, exactly \textit{what} is to be
transferred. Consequently IoT applications today are often restricted
to specific domains.  A promising approach for data-layer integration
is the idea to weave the IoT with the \textit{Semantic Web}.  The
Semantic Web provides the tools and language to describe the real
world using the Resource Description Framework
(RDF)~\cite{citeulike:99942}, a standard for encoding knowledge in a
universal way that is extensible, machine-readable and can connect
facts across different locations.  It also is the basis of the Linked
Data Cloud~\cite{sweo2007linking}, where data from different sources
in the Web is fused to become a unified and huge data set describing
the world as we know it.

Most application continue to express data on the embedded
device in a specific, limited vocabulary and adapt it to the Semantic Web
standards only at a preceding proxy.  This leads to a number of limitations:
Embedded devices continue to be simple providers of sensor data, thus not
allowing any direct exchange between embedded devices of different proxy
domains. Furthermore, it makes it impossible to take a device operating in a network running
some application and move it to another network running a different
application. This is acceptable in today's domain-specific
applications, but not for the massive IoT networks we will see in the
near future, installed and maintained by non-experts, who expect
devices to be general-purpose and to be cross-compatible, just like
any other computing device they are using. This is only achievable if
devices are self-contained and independent of a specific
infrastructure, requiring them to generate and process data without
proxy components.

% not in the context of an 

% This limits the set of applications that can make use of the device
% to those specifically written for the particular device and applications that
% utilize the device through a proxy, universally applicable embedded
% applications are not possible.  That is because it might be equipped with sensors that have not
% been foreseen by the foreign installation, uses other units of measurement, or
% provide a different level of data abstraction.
% We argue that it is thus beneficial to store RDF directly on the embedded
% devices, to make them self-contained by holding universally understandable
% self-descriptions and observation encodings.

%Embedded devices with this capability advance from generators of data that needs
%interpretation to providers of (machine-)understandable facts, that can be
%combined with facts from other sources and be reasoned upon.

This paper introduces the \TS, a flexible and portable database for efficient storage of
RDF on embedded devices. It allows to quickly add RDF-processing
capabilities to an application. The \TS\ runs on many
different platforms (including Contiki, TinyOS, Arduino/Wiring,
Android, and iOS). It utilizes Flash memory where it is present,
but can also store data in RAM when required. This enables
applications that leverage the huge data sources available in the
Linked Data Cloud, and that are no longer restricted by
domain-specific languages (and ``data silo'' issues). We also show
that there is little overhead to be paid for the generality, both
energy-~and storage-wise.

The \TS\ is part of the SPITFIRE architecture~\cite{Pfisterer2011},
which includes 6LoWPAN, CoAP, RDF compression, and more, all available
as platform-independent Wiselib components. This provides an easy way
of building applications that use fused knowledge from the Semantic
Web and embedded devices.
%
%different components~\cite{Pfisterer2011} that
%allow embedded devices to be self-describing and to
%integrate them into the Semantic Web and provides a number
%of tools and applications to exploit this universality.

%a flexible system holding tuples of any type of data, which is
%part of this architecture.
%We demonstrate
%its use for managing RDF data sets efficiently on constrained embedded devices.
%Using the Wiselib~\cite{bcfkkp-wgalhsn-10}, a platform-independent algorithms library for embedded  devices, it runs on a multitude of platforms and operating systems (including Contiki, TinyOS, Arduino, iSense, OpenWRT, and many more). It is modular
%by design, allowing an application to find the perfect balance between data structure capabilities and resource efficiency. The Wiselib principles allow the TupleStore to offer this flexibility at virtually no cost for the resulting compiled image
%in terms of code size.

The rest of this paper is structured as follows: In
Chapter~\ref{sec:problem_statement} we will introduce the challenges
accompanying our idea and discuss related work. In Chapter~\ref{sec:architecture}
we present the architecture of our system before we provide details about our
compression methods in Chapter~\ref{sec:compression}. Finally in
Chapter~\ref{sec:evaluation} we compare our solution to the well-known
embedded tuple stores Antelope and TeenyLIME and then conclude in
Chapter~\ref{sec:conclusion}.

%%% Local Variables: 
%%% mode: latex
%%% TeX-master: "article"
%%% End: 

\section{Problem Statement and Related Work}
\label{sec:problem_statement}

%We consider the problem of semantically describing embedded devices and their
%surroundings. More specifically we investigate the feasibility of storing and
%processing semantic descriptions adhering to the \textit{Resource Description
%Framework} (RDF) on resource-constrained embedded devices.

Devices in the IoT are diverse and heterogeneous in several aspects. The IoT
vision strives to integrate these devices not only with each other but also
with the current Internet in order to create a universal web of knowledge
covering both static data and live observations.
%
% - Hardware
Depending on vendor, application purpose and deployment context devices may
differ vastly in terms of utilized hardware. Thus a variety of combinations of
processor types, available memory, means of communication and energy
constraints is encountered in the field.
% - OS
Additionally, a variety of operating systems, middlewares and applications for
these devices are already in daily use. Some of them support multiple devices,
Contiki~\cite{1367266} and TinyOS~\cite{springerlink:tinyos}
%and --- targeting the more powerful systems --- several derivations
%Linux~\footnote{\url{http://www.kernel.org}}
being probably the most prominent representatives of this class.
%Still they focus on a very specific set of platforms by assuming certain
%restrictions in the available hardware inherently in their design.
%Especially in the more resource-constrained embedded device section
We also encounter proprietary systems specifically designed for a rather
limited set of hardware platforms.  In addition to this diversity in terms of
hardware and software we observe varying deployment contexts and usage
scenarios of these devices.

We raise the question of how these devices can be integrated in a meaningful
way such that knowledge can be shared across varieties hardware, software or
application contexts in order to allow universal auto-configuration and a
style of application development that can make use of data located in the Web as
easily as live sensor descriptions.
This breaks down into several sub-problems:

\subsection{Component Portability} The plethora of available devices has
led to a patchwork of individual solutions for each device, tailored to the
system specifications. Contiki and
TinyOS have been developed as embedded operating systems which run on several
hardware platforms to ease up this situation. Still, software components
 often are only available for one OS. For the application developer,
 porting code to all target platforms and maintaining parallel software
 lines is required.
Because of this fact, we use the
Wiselib~\cite{bcfkkp-wgalhsn-10} as foundation for the \TS{}.
%already cited in intro\cite{bcfkkp-wgalhsn-10}
The Wiselib is a platform-independent algorithms library for embedded devices,
running on a multitude of platforms and operating systems (including Contiki,
TinyOS, Arduino, iSense, OpenWRT, and many more). It is modular by design,
allowing an application to find the perfect balance between data structure
capabilities and resource efficiency.
%The Wiselib principles allow the TupleStore to offer this flexibility at
%virtually no cost for the resulting compiled image in terms of code size.  To
%our knowledge, the Wiselib is the only algorithms library so far providing
%platform-independent algorithms and data structures for embedded systems.
By building on the Wiselib, the \TS{} can be compiled and optimized for a wide
range of platforms, making it widely available for future applications.

\subsection{Standardized and Open Data Layer}
% Given two devices that agree on protocols and data container format, a
% successful data exchange can not be guaranteed, as the container format
% itself only provides the means to structure data but does not define
% which structure to apply and what data to communicate.
Even given physical interoperability, common communication protocols and data
serialization formats, applications on different devices can not necessarily
exchange data successfully. While serialization formats provide
structure for data, content and semantics stay undefined and
thus different for every application.  Several standardization efforts have
been made to agree on a common data layer: The \textit{Open Machine Type
Communication Platform} (OpenMTC)~\cite{openmtc}
%~\footnote{\url{http:///open-mtc.org/index.html}}
provides machine-to-machine (M2M) communication methods on top of several
standardized protocols and provides features such as service discovery,
routing and notification.  OpenMTC defines a set of XML schemas that can
describe sensors and actuators in several aspects. Although exhaustive, this
approach is limited to a certain descriptive domain and does only provide
user-given device annotations via tags with limited descriptive capabilities.

%Several approaches to connect the Semantic Web with the IoT have been
%proposed in the past years: On the semantics side this challenge has risen
%the question on how to account for the fact that information in sensing
%environments has a much stronger dependence on time and location than former
%RDF documents~\cite{Janowicz2010,Kessler2010}.
%%~\cite{Janowicz2010,Kessler2010,Leggieri2010,Kyzirakos2010,Barnaghi2010,Tran2010}.
%Other work concentrates on how to describe the connections and interactions
%of semantic data sources and processors, to extend the formerly relatively
%static semantic world by models for changing
%state~\cite{Le-Phuoc2010,Whitehouse2006}.
How to convert the often compactly and proprietary communicated sensor data to
semantic documents has been answered by a multitude of proxy-oriented
approaches~\cite{Huang2008,Song2010,Le-phuoc,Hasemann2013}.  The \textit{Open
Geospatial Consortium} (OGC) published the \textit{Sensor Observation Service}
(SOS)~\footnote{\url{http://opengeospatial.org/standards/sos}}, a standard for
web service interfaces to sensor networks, defining a set of XML schemas
describing sensor meta data, observations or geography.  As an extension to
SOS, Henson et al. introduced \textit{SemSOS}~\cite{Henson2009}, proving a
bridge to the Semantic Web and allowing advanced semantic queries and semantic
reasoning.  \textit{Internet Connected Objects for Reconfigurable Ecosystems}
(iCore)~\footnote{\url{http://www.iot-icore.eu}} is a project funded by the
European Union that proposes such a proxy-based framework that composes
semantic descriptions of embedded devices into semantic descriptions of
observed real-world objects~\cite{kelaidonis2013cognitive}.  All of these have
in common that the exposed data will not carry universal semantics before
being transformed by the proxy system, making the device inherently depending
on its proxy to be useful.
%A variety of approaches have been published focusing on the idea of a smart
%gateway-like component that translates between a restricted format for
%communication with the embedded network and a more universal representation
%such as RDF to the ``outside'' internet.
%

\subsection{Independence of Translating Infrastructure}
%The problem of describing embedded devices, e.g. sensor networks semantically
%has been considered in the past, especially using Semantic Web technologies.
%
%\subsection{The Resource Description Framework and the Semantic Web}
%
% Some RW copy-paste from dcoss
%
In order for embedded devices to be useful independently of a supporting
infrastructure and thus utilize the semantic data on the device for universal
auto-configuration and communication between embedded devices, it is necessary
to manage a certain amount of data on the device itself.  Sadler and Martonosi
have developed a database for embedded devices that is tailored for usage in
Delay Tolerant Networks (DTN)~\cite{Sadler2007}.  Distributed tuple spaces
have been presented in the TeenyLIME~\cite{Costa2007,Ceriotti2011} and
Agilla~\cite{Fok2009} systems, which are available for TinyOS.  Recently,
Tsiftes et al.~\cite{Tsiftes} have introduced a relational database for
devices running the Contiki operating system.  All of these allow managing
general-purpose data, however do not feature compact representation of RDF
data.  As RDF in its common serializations (such as RDF/XML and N3) tends to
be quite verbose, efficient compressing storage schemes are crucial to employ
semantically self-describing embedded devices.

\subsection{Resource Constraints.} When storing data on embedded devices, one
has to overcome strong resource constraints. Many sensor networks are battery
powered, making energy an extremely critical resource.  Therefore a storage
system has to ensure that operations to store, delete or access the data will
not result in excessive battery drain. Other systems might be equipped with a
considerably restricted amount of memory, sometimes even 
shared with operating system and program code.
Therefore, a database system which aims to store RDF data on embedded devices
profits from lightweight compression techniques for saving both energy and
space.
Compression of RDF data has recently been addressed by Fern\'{a}ndez et
al.~\cite{Fernandez2010}, with the \textit{Header-Dictionary-Triples} approach
that compresses tuple elements into a dictionary, followed by a compressed
representation of triples of dictionary keys.  This approach however can only
be successfully applied to large data sets and focusses on serialization in
the sense that it is not well suited for updating the compressed RDF on
tuple-level.  Su and Riekki~\cite{Su2010} proposed an approach for producing
RDF on embedded devices by the use of templates, that is, a predefined set of
statements containing place holders to be filled in with e.g. current sensor
values.  This approach however assumes a known structure to the semantic data
that is to be expressed which contradicts our demand on the ability of
expressing arbitrary and possibly unforeseen facts.

%%% Local Variables: 
%%% mode: latex
%%% TeX-master: "article"
%%% End: 

\section{Architecture}
\label{sec:architecture}

%General TS, interface, very short overview over compnoents, tuples.

%% --- Chapter structure

%In order to provide  RDF descriptions on resource constrained devices, a
%number of components must play together. This paper focusses on the Wiselib
%Tuplestore, which we consider the core component of such an architecture, in
%order to better understand its rule, we will briefly discuss interactions with
%other components in this chapter.
%In order to ascertain a high degree of modularity and flexibility in terms of
%available resources, the Wiselib TupleStore itself is split into a number of
%subcomponents. These can be parameterized, exchanged and combined in several
%ways in order to adapt the Wiselib TupleStore and will also be discussed.
%Finally, we will provide an application scenario that makes use of the
%flexibility provided by the Wiselib TupleStore.

% --- Why do we need modularity etc....

When designing a software architecture for the IoT, a multitude of different
platforms has to be considered. These platforms not only differ in processing
speed, but also vastly wrt.\ memory size and available energy.
%Note that the Wiselib TupleStore is designed to hold all data in the nodes
%RAM so that it can be used on nodes that do not feature additional flash
%memory such as the TelosB or MicaZ for user data.  Depending on the chosen
%platform, the code size might or might not be restricted independently from
%the usable RAM at runtime.
In an IoT network, an application is likely developed under constraints on
available energy and the amount of data it must be able to store and/or
process. If the task at hand involves a certain amount of communication, it
might be worthwhile invest some energy for compressing data. For applications
that do not transmit a lot of data, the maintainer might decide to avoid the
energy cost of an encoding mechanism altogether and reduce the code size
instead.
In order to satisfy these high demands on flexibility, a modular architecture
is essential. In this section, we will illustrate the different components of
our architecture and show how the choice of their assembly allows trade-offs
between memory footprint, code size and energy consumption.
%An overview of the the components of the Wiselib TupleStore is given in
%Figure~\ref{fig:tuplestore_architecture}.

\subsection{The \SPITFIRE Software Stack}

\begin{figure}
	\centering
	\def\svgwidth{.5\textwidth}
	\footnotesize
	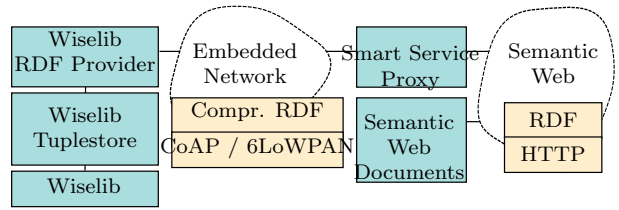
	\caption{Simplified view of the SPITFIRE architecture: The Wiselib RDF
Provider and the \TS{} provide RDF on tuple- and document level on the embedded
device. The Smart Service Proxy then connects the embedded network to the Semantic
Web.}
	\label{fig:software-arch}
\end{figure}

In our vision of the IoT, knowledge on the Web as well as (live) descriptions of
embedded sensing devices are encoded in RDF and interlinked with each other
across physical domains.
The SPITFIRE project realizes this vision with a stack of components shown in
Figure~\ref{fig:software-arch}: The foundation for managing RDF data on the
embedded device is the \TS{} allowing the device to work with RDF
data on a tuple level. On top of that, the Wiselib RDF Provider~\cite{rdfprovider} enables a
document view, a notification mechanism and several pluggable RDF
serializations and communication protocols.
These communication mechanisms can be utilized to access
data in the Wiselib TupleStore electively on a tuple- or document level.
Moreover the communication interface allows connecting the embedded device to
a \textit{Smart Service Proxy}~\cite{Hasemann2013} instance which can expose
the descriptions of the device to the Semantic Web and allows a user to query
the embedded devices using SPARQL~\cite{SPARQL} over standard web services, providing for
caching, push/pull mechanisms, format translation and several other features.

\subsection{TupleStore Functions}

We designed the TupleStore as a flexible and lightweight data storage that
provides the following operations: \texttt{insert} inserts a new tuple into
the store, \texttt{query} finds tuples matching a tuple template with
wildcards and \texttt{erase} erases a given tuple from the tuple store.
%
%\begin{description} \item[\texttt{insert}] Insert a tuple into the store (if
%not yet existent) \item[\texttt{query}] Given a tuple template containing an
%arbitrary number of wild cards return all matching tuples.
%\item[\texttt{erase}] Erase a tuple from the store.  \end{description}
%
For enabling lightweight configurations, the \TS{} core offers only restricted
means of querying through the \texttt{query} method described above.  The
SPITFIRE architecture provides two components on top of the \TS{} that allow a
more sophisticated tuple selection: The Wiselib RDF Provider provides the view
of (potentially overlapping) documents, that is, sets of tuples.  Additionally,
SPITFIRE offers also an in-network query processing mechanism which also
includes a local component that can be used for more complex queries.

%Note that the TupleStore does not feature a complex query parser comparable to
%SQL and the like as we believe in most applications for embedded systems
%all complex queries are known at compile time and thus can be expressed as
%specialized code that utilizes the \textit{query} function of the TupleStore.
%Note that although we assume a priori knowledge about the stored data, by
%the use of RDF we always know the ``form'' of the data at programming time.
%That is, unlike relational databases we do not have to cope with different
%table formats etc..
%A system that allows more complex and also distributed queries is
%work in progress.

\subsection{Components}

\begin{figure}
	\centering
    \def\svgwidth{\columnwidth}
	\footnotesize
	%\tiny
	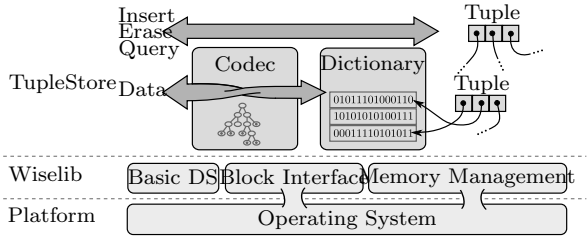
  \caption{\TS{} architecture. Tuples hold keys pointing
    into the optional dictionary, which in turn holds optionally compressed elements.}
  \label{fig:tuplestore_architecture}
\end{figure}

In order to implement the TupleStore in a modular and resource-efficient
manner, it is crucial to reduce the cost of modularity to a minimum.
Mechanisms such as runtime dispatch of method calls (as in C++'s virtual
inheritance mechanism) are undesirable, as they add overhead at runtime (and
thus energy consumption) and code size, even if it is known at compile-time
which modules are to be selected. Also, such mechanisms create optimization
barriers, i.e.\ disallow the compiler to apply optimizations such as method
inlining.  We rely on the modularity approach of the Wiselib: Components are
implemented as templated C++ classes, called \textit{models}. They are
instantiated by the compiler, receiving operating system specifics as template
parameters.  This way, a component that is not being used will never be
instantiated, i.e.\ the component will not use any code space. This allows for
a high degree of modularity without code size or runtime penalties.  As
typical in the Wiselib, we use \textit{concepts} to define the interfaces of
our components. Concepts describe properties of models.  We briefly introduce
our components, some of them grouped together for brevity. A conceptual
overview of the interactions of the components is given in
Figure~\ref{fig:tuplestore_architecture}.

\subsubsection{Tuples} Depending on the source of the data that is to be
inserted into the TupleStore, different internal tuple representations might
be appropriate. E.g., certain elements might be generated on demand or be the
same for a huge number of tuples. For this reason, TupleStore components do
not force their users to use a specific tuple implementation but provide
templated methods that accept any type adhering to the \texttt{Tuple} concept.
This concept requires anything that is to be accepted as tuple to implement
the following methods: \begin{description} \item[\texttt{size}] Report the
size (number of elements) of the tuple.
%\item[is\_wildcard] Return true iff the element at the given position is a
%wildcard (i.e. should compare equal to all other elements when used in a
%TupleStores \texttt{find} method).
\item[\texttt{access}] Access a given tuple element (read/write).
\end{description}

%Components, optionality
%\subsubsection{Container}
%Most wiselib/STL containers can be used
%\subsubsection{Dictionary}
%Role of dictionary, implementations
%\subsubsection{Codec}
%Huffman

\begin{figure}
	\centering
	\def\svgwidth{1.0\columnwidth}
	\footnotesize
	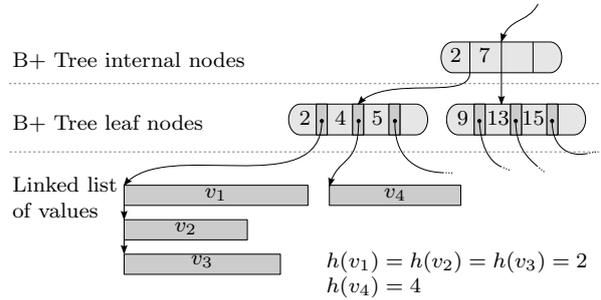
	\caption{B+ tree based hash set used for dictionary and tuple container
implementations.
Hash values of the inserted tuples (tuple container) / string values
(dictionary) are used as keys. In case of a hash value collision, a linked
list of values is maintained. For the dictionary, memory addresses of the
strings are used as lookup key for $O(1)$ access.
}
	\label{fig:hbptree}
\end{figure}

\subsubsection{TupleStore} The TupleStore component provides the operations
\texttt{insert}, \texttt{erase} and \texttt{query} using a container model to
hold the tuples as well as a selectable dictionary implementation for holding
the tuple elements.  Depending on the memory management capabilities of the
platform, a user would typically use either a static vector or a dynamic
linked list as tuple container, however other containers such as dynamic
vectors or set implementations are also possible.
In addition to these RAM-based data containers, the user may also decide to
store tuples on a block device such as an SD cards, using a container
implemented as a external hash set based on the B+ tree~\cite{btree}, see
Figure~\ref{fig:hbptree} for illustration.
%The Wiselib provides a variety of modules that implement the block concept,
%beyond others, SD card abstractions fill this use a linked list, although
%other containers such as static or dynamic arrays are also possible.
Internally, the actual stored data %(e.g.\ strings) will be substituted with
the much more space-efficient dictionary keys that point to corresponding
entries. The TupleStore will ensure that on read access, the element values
are fetched from the dictionaries transparently such that the user is not
concerned with the dictionary mechanism.

It is possible to configure (at compile time), which elements of a tuple
will be managed using the given dictionary, which enables avoiding
unnecessary dictionary operations on value types for which they are not profitable
(e.g.\ integers). If the code size and/or runtime overhead for dictionary
compression is not considered reasonable, it can be switched off altogether,
yielding a TupleStore that stores tuples in its container ``as-is''.

%<RDF>
%In the Semantic Web context this provides a
%lightweight representation of the semantic graph that is independent
%from the actual element data. Since there is a one-to-one
%correspondence between dictionary keys and values, this structure
%would allow for reasoning with only occasional or no dictionary
%lookups (e.g. for communication with others).
%</RDF>

\subsubsection{Dictionary} The task of a dictionary is the efficient storage
of tuple element data.  Dictionaries provide methods for inserting data
(returning a corresponding key), for data access by key and for data deletion.
A dictionary can, in several ways, exploit redundancies in the stored data in
order to reduce memory usage.  We implemented three dictionary models: The
\textit{AVL Dictionary}, the \textit{Prescilla Dictionary} and the
\textit{Chopper Dictionary}, all of which compress data by storing for each
element a count such that elements occurring multiple times only need to be
stored once.  Similarly to the tuple container, we also provide a
block-storage based dictionary implementation based on the B+ hash set
described in Figure~\ref{fig:hbptree} which can be exchanged transparently
with the RAM oriented dictionaries.  See Section~\ref{sec:compression} for an
in-depth discussion on the compression schemes used in the dictionaries we
implemented.

\subsubsection{Codec}
Given a codec, the \TS{}
transparently encodes/decodes data (i.e.\ tuple elements), such that internally
only compressed/encoded data is stored while the user can work with plain text
data and does not need to be concerned with the encoding process.
It is however also possible to directly access the encoded data such that it
can be transferred to other nodes without the need for a decoding operation.
In
Section~\ref{sec:huffman} we will introduce in detail our Huffman codec
that compresses data using a predefined Huffman tree.  In
contrast to the dictionary approach, a codec does not store any tuple
elements but rather transforms them between plain and
encoded format.

%\subsection{Network Applications}

%\todo[inline]{A) huffman-kodiert zeug rumschicken}
%\todo[inline]{B) SHDT, INQP koennen direkt dict-keys speichern}
%This element-wise compression has some interesting properties: First, it
%provides additional compression that plays
%together nicely with dictionary compression while still
%being usable independently of it.
%Second, unlike dictionary compression, a codec is independent
%from the state of the TupleStore and thus two different nodes
%compressing the same elements will always yield the same results. Thus, given a
%communication partner with a common codec, this allows to
%exchange encoded tuples, while saving energy during radio
%transmissions by transferring less data.
%Even if we consider a communication partner that does not have the according codec
%compiled-in (e.g.\ for saving code size),
%it will still be possible for it to store, compare and pass on the tuple data
%to other nodes.
%We can even go as far as to consider a scenario in which almost all
%nodes only ever work with compressed tuple elements in-network and
%only a few border nodes connected to outside networks
%need to ``translate'' (i.e.\ compress/decompress) data.

%A detailed analysis of the possible trade-offs between code size, RAM
%usage and energy consumption for different configurations can be found
%in Section~\ref{sec:evaluation}.

%- Ref SHDT: Application for dictionary
%- Encoded network

\section{Compression Components}\label{sec:compression}

RDF data consists of \SPO triples, whereas the triple
elements are either URIs, literals or local identifiers.
As identifiers, URIs and string literals usually have a relation to natural
language, their symbols are likely to be unequally distributed, that is, some
symbols are more likely to be encountered than others.
Furthermore, few subjects (be they referring to abstract concepts or real-world objects)
can be described sufficiently with a single triple, that is, repetition of
elements is to be expected in realistic RDF data.
When elements share a (semantic) domain in the sense that they are related in
meaning, it is likely the case that they also share common URI prefixes.
These different types of redundancy can be efficiently exploited by a combination of
element-wise and cross-element compression mechanisms which we present in the
following sections.

\subsubsection*{Huffman Coding}\label{sec:huffman}

A standard approach for string compression is {\em Huffman
coding}~\cite{Huffman1952}, a variable-length encoding that depends on the
distribution of plain text symbols.  The codec is represented in form of a
binary tree with plain text symbols in the leaves. For each of these plain
text symbols, the unique path to the root of the tree yields the corresponding
code symbol. The tree is formed by arranging all possible plain text symbols
according to their frequencies in the data such that the most frequent plain
text symbols are encoded with the shortest codes.  It has been shown that the
application of Huffman coding can lead to significant energy savings as the
size of transmitted messages can be reduced. See for example the works of Yeo
et al.~\cite{Yeo2009} or Yuanbin et al. \cite{Yuanbin2011}.
%Surprisingly, little research has been conducted to code size and memory
%efficient storage concepts for full Huffman trees.  Reinhardt et
%al.~\cite{Reinhardt2010} avoid this problem by storing only  the encoding for
%the $n$ most frequent characters while encoding the others with standard
%ASCII\@. This limits the tree size reasonably but comes at the cost of
%compression gain.
Our approach uses \textit{succinct trees}~\cite{Munro1997a} to store the
Huffman tree with high space efficiency. Succinct data structures focus on
storing data with a space efficiency, as close as possible to the theoretical
limits.  Arroyuelo et al.~\cite{Arroyuelo} give a detailed overview about
available succinct tree implementations in their work.  Our codec manages a
compact Huffman tree in 416 bytes of storage that is defined for the whole
ASCII alphabet and can thus be used to compress and decompress any tuple
element.

\subsubsection*{AVL Dictionary}\label{sec:avl}

For storing repeated tuple elements (be they strings, integers or any other
data type), we add reference counting to the stored elements.
To find the element record and the associated count,
we need a data structure to keep track of all those records. By using the well-known approach
of AVL trees \cite{avl}, we obtain a dictionary that can
guarantee to do insert and delete operations in $O(\log n)$ element comparisons,
where $n$ is the number of inserted elements.
By using the memory addresses of nodes holding element values as
dictionary keys, we can additionally guarantee a value lookup by key in
$O(1)$.

\subsubsection*{Prescilla Dictionary}\label{sec:prescilla}

As RDF data tends to have a high amount of common prefixes, element
comparisons in the AVL tree repeatedly compare the same prefixes in different
elements, thus making the runtime also dependant on the average element
length.  The \textit{Prescilla Dictionary} utilizes a variant of the radix
tree (or PATRICIA tree), a trie data structure originally introduced for text
indexing~\cite{Morrison1968a}. The constructed tree does not hold complete
elements in its nodes but substrings such that the concatenation of node
values along a path from a root to a leaf forms a string present in the
dictionary.  Our data structure contains several optimizations over the radix
tree to make it more efficient for use on resource-constrained embedded
devices.
Like with the AVL Dictionary, by using the memory addresses of nodes holding
element values as dictionary keys, we can additionally guarantee a value
lookup by key in $O(1)$.

\subsubsection*{Chopper Dictionary}\label{sec:chopper}

Both the AVL dictionary and the Prescilla dictionary provide a tree structure
holding variable-sized data in the nodes. Thus, in order to create a new node,
a sufficient amount of memory must be allocated and freed again when the node
is not needed anymore. This kind of memory management is available on some
platforms either in hardware or software, for example the iSense operating
system supports \texttt{malloc()} and \texttt{free()} in the usual libc style.
Some platforms, such as Contiki (e.g. on the TMote Sky) do not feature such a
mechanism. The Wiselib does provide memory allocators to compensate for this,
however, software allocators naturally come with a certain overhead in terms
of CPU usage (for allocation) and memory usage (for supporting data structures
and fragmentation).  In order to address this issue, we provide the
\textit{Chopper Dictionary} which does not require dynamic memory allocation.
The key idea is to only handle fixed-size chunks of strings in a statically
allocated table. In order to connect the different chunks to a complete
string, special \textit{meta chunks} are inserted that do not contain string
data but references to string chunks in order to encode longer strings.  While
this approach can in general not exploit all common-prefix redundancy present
in the data due to the fixed-size string partitions, its very low overhead
(one additional byte per chunk, no dynamic allocation needed), make it a
useful asset for heavily resource-constrained devices.
%it has virtually no requirements whatsoever on the underlying platform and
%can still exploit a large fraction of redundancies across strings.

\section{Evaluation}
\label{sec:evaluation}

In this chapter we investigate the features and performance properties of
the \TS{} in comparison with existing embedded databases Antelope and TeenyLIME.
While we are convinced that these approaches are
impeccable for storage of short fixed-length data, such as numeric values, we
hope to illustrate that for the storage of RDF, the \TS{} offers
certain advantages.

First, we will briefly discuss the datasets we considered and the
effectiveness of our compression mechanisms on them. We then discuss the
modularity and code size of different \TS{} configurations. Finally, we evaluate
execution times and energy consumption of the \TS{} operations and compare them
practically to TeenyLIME and Antelope.

%\subsection{Properties of RDF data}
%\label{sec:rdf_properties}

%RDF data exhibits several redundancies, which partly stem from its general
%structure (repetition of elements, use of URIs) and partly from properties of
%the content described with RDF (e.g. the amount of URIs of a common domain).
%In a preliminary study we thus analyzed the properties of three different RDF
%data sets:
\subsection{Datasets}
\label{sec:datasets}

In order to address the diversity of RDF descriptions found in different
application contexts, we consider three different datasets for analysis of our
compression components:

%\begin{description} \textbf{BTCSAMPLE.}
\paragraph{BTCSAMPLE} This dataset is a random selection of 4796 triples from
the Billion Triple Challenge (BTC)
2011\footnote{\url{http://km.aifb.kit.edu/projects/btc-2011/}}, with an
uncompressed file size of 1040kB.  As the Billion Triple Challenge targets at
collecting large numbers of triples, this excerpt is very diverse in terms of
contents.  We thus expect very few repetitions of tuple elements and also few
shared prefixes, in this regard the BTCSAMPLE dataset will provide a hard
compressible RDF dataset.

\paragraph{SSP} The Smart Service Proxy (SSP) dataset was assembled using the
output of an instance of the \textit{Smart Service
Proxy}\footnote{\url{https://github.com/ict-spitfire/smart-service-proxy}}, a
software system that converts the output of over 300 real sensors into
semantic descriptions.  The output consists of 4859 triples and has a total
size of 883kB. Most of the URIs share as a prefix the URI of the service,
which we expect to be beneficial for compression.
%This is a very realistic set of data when considering an IoT application as
%it is a collection of semantic descriptions of sensors and measured values
%which perfectly fits the kind of data an application on an embedded IoT
%network would usually deal with. 

\paragraph{NODE} The \textit{NODE} dataset contains a typical RDF description
of a single sensor node.  It describes a temperature sensor with its measured
value, unit of measurement, value range, and method of measurement (stimulus).
The dataset has been generated using the publicly available LD4Sensors web
application%
\footnote{\url{http://spitfire-project.eu/incontextsensing/ld4sensors.php}}.
The generated RDF file holds 73 triples and has a size of 7.6kB.  Due to its
suitable content and size, we will conduct the evaluations on the TMote Sky
devices (see next chapter) using this dataset.
%In contrast to the SSP dataset described above, NODE is less homogeneous as
%it uses more external references and also describes e.g.\ the physical
%effects exploited for measurement.  \end{description}

\begin{figure}
	\centering
	\includegraphics[width=.35\textwidth]{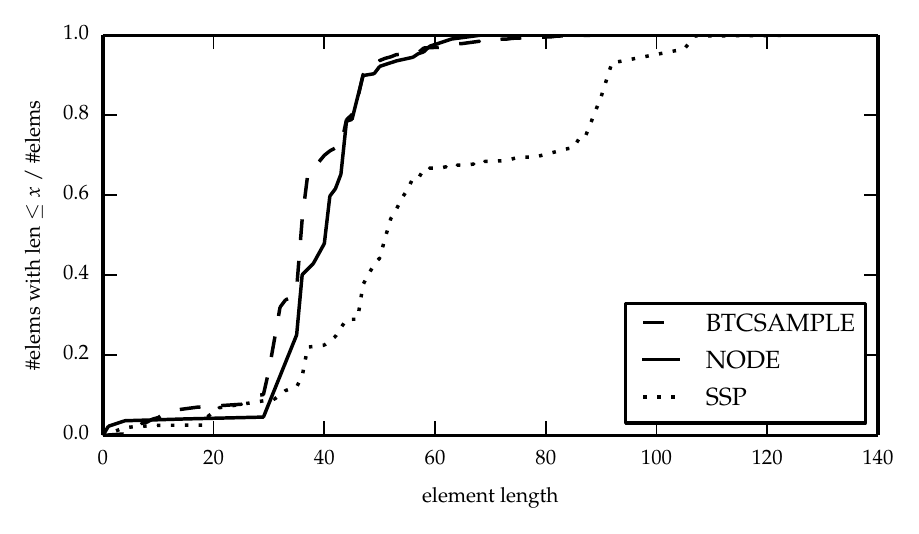}
	\caption{Distribution of string lengths (bytes) of RDF elements for our datasets.}
	\label{fig:lengths}
\end{figure}

%Figure~\ref{fig:lengths} shows the distribution of element lengths of the data
%sets we considered. While e.g. 50 bytes would be enough to store 90 percent of the
%elements of the NODE and BTCSAMPLE data sets, they would only be sufficient
%for less than 40 percent of the SSP data set. Even within one data set, a
%fixed-length string storage is bound to inefficiently use the available
%storage space, as the ``plateaus'' suggest that only very few elements would
%fully utilize the reserved string size.

%\begin{figure}
	%\centering
	%\includegraphics[width=.35\textwidth]{fig/prefixes.pdf}
	%\caption{Distribution of common prefix lengths (bits) among different
%datasets, exploitable eg. by the Prescilla Dictionary.}
	%\label{fig:prefixes}
%\end{figure}
%\todo[inline]{write oder beide rdf figures raus}

\subsection{Compression}

\begin{figure*}
  \centering
  \subfloat{
    \centering
    \includegraphics[width=.35\textwidth]{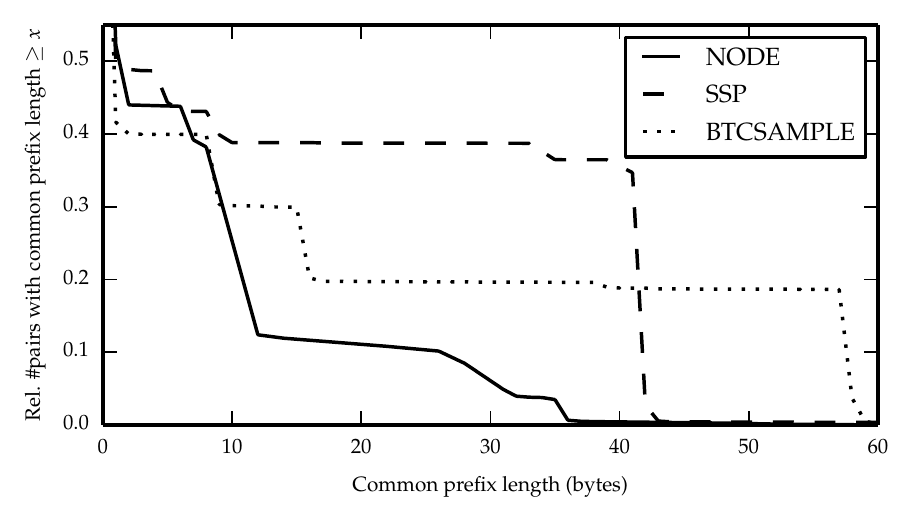}%
	%\caption{Common prefixes}%
    %\label{prefixes}%
  }%
  %\end{subfigure}
  %\begin{subfigure}[b]{.3\textwidth}%
  \subfloat{
    \centering
    \includegraphics[width=.345\textwidth]{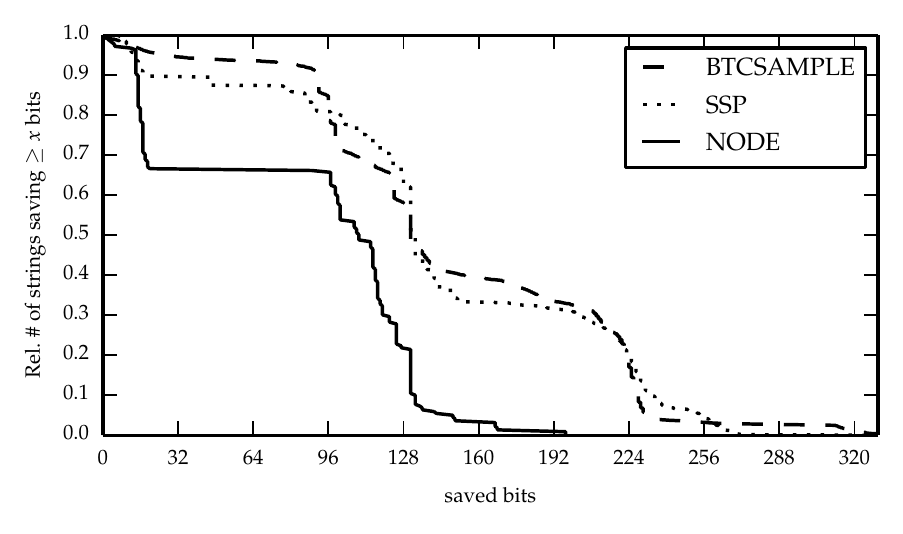}%
	%\caption{Huffman compression gains}%
    %\label{huffman}%
	}
  %\end{subfigure}
  %\begin{subfigure}[b]{.3\textwidth}
  \subfloat{
    \centering
	%\begin{center}
	\raisebox{1.25cm}{%
	\footnotesize
	\begin{tabular}{|l|c|c|c|}
		\hline
		 & BTC & SSP & NODE \\
		\hline
		Original & 787k & 880k & 7.3k \\
		Huffman & 514k & 615k & 5.0k \\
		Huff. \& AVL & 764k & 265k & 5.5k \\
		Huff. \& Prescilla & 689k & 242k & 5.8k \\
		\hline
	\end{tabular}}
	%\caption{Total compression gains}
	%\label{gains}
	%\end{center}
  }
  %\end{subfigure}
  \vspace*{-1em}
	\caption{Left: CDF of common prefix lengths across all pairs of elements for each data
set (higher is better compressible by dictionaries). Center: CDF of saved bits when
compressing element-wise using the Huffman codec (higher is better
compressible by Huffman). Right: Overview of our datasets in compressed form.}
	\label{fig:compression}
\end{figure*}

Figure~\ref{fig:compression} shows the effectiveness of our different
compression approaches: As the prefix commonalities differ notably between
different datasets, while the effect of Huffman compression is much more
independent from the chosen dataset and reliably saves about 30 to 35 percent
storage space.  Adding a Prescilla dictionary drastically changes the picture
(Fig.~\ref{fig:compression}c): For BTC and NODE the compression ratio has
decreased due to overhead and the lack of many common prefixes, while for SSP
72.5 percent of the data can be compressed.

\subsection{Heterogeneity and Code Size}
\label{sec:codesize}

\begin{table*}
	\begin{center}
		\begin{tabular}{|c|c||c|c|c|c|c||c|}
			\hline
			Platform & \multicolumn{4}{|c|}{TS \& Dictionary} &  \multicolumn{2}{|c||}{CodecTS \& Dictionary} & Antelope \\
			\hline
			& TS only & Prescilla & AVL & Block & Prescilla & AVL & DB Kernel \\
			\hline

			iSense 5139    &  3332 /    32    &  5780 /    36    &  8948 /    40  & -               &  9960 /    36    & 13184 /    40    & - \\
			iSense 5148    &  1828 /    28    &  3444 /    36    &  5112 /    40  & 13600 / 11072   &  5840 /    40    &  7568 /    40    & - \\
			Contiki/MicaZ  &  3490 /    13    &  5910 /    16    &  8562 /    18  & -               &  9462 /   433    & 12120 /   435    & \multirow{2}{*}{3570 / ?  \cite{Tsiftes}}   \\
			Contiki/TelosB &  1634 /    14    &  3710 /    16    &  5022 /    18  & -               &  5962 /    18    &  7282 /    20    &  \\
			TinyOS/TelosB  &  1526 /    14    &  3534 /    16    &  4786 /    18  & -               &  5792 /    18    &  7050 /    20    & - \\
			TinyOS/MicaZ   &  3490 /    13    &  5910 /    16    &  8544 /    18  & -               &  9462 /   433    & 12102 /   435    & - \\

			\hline
		\end{tabular}
	\end{center}
	\vspace*{-.5em}
	\caption{code sizes for various platforms and
		  configurations. {\small (The notation is ``ROM / RAM''.
	For example, on TelosB we can provide a TupleStore featuring 
	\it insert\rm, \it delete\rm, and \it query \rm in 1.6kB of code
	size.}}
	\label{fig:codesize}
\end{table*}

The \TS{} offers a multitude of configurations which allow different trade-offs
in terms of code size, RAM usage and energy consumption.  On the lower end of
the code size scale, we have a feature-minimal TupleStore configuration that
still allows insert, deletion and querying of tuples on multiple platforms.
%To create an impression on how the different components can be used to adjust
%the code size/RAM/energy trade-off, we analyze a number of TupleStore
%configurations. For each of these configurations we ran experiments on code
%size, heap usage and energy consumption.
For code size considerations, we compiled the same source code for different
platforms (see Table~\ref{fig:codesize}) and calculated the difference in size
to an empty Wiselib application in order to obtain the code space consumption
of the individual component. As it is not possible to instantiate all
components usefully standalone, we take a cumulative view on the components,
i.e.\ add them up one after another and/or exchange them so we can reason
about the code size difference of that particular step.

%For analysis of the memory footprint, we compiled the TupleStore for iSense
%5148 nodes and continuously inserted tuples from our datasets, monitoring the
%available memory. For this particular platform, code and heap share
%a memory area thus the amount of free memory at any point in time does not
%only depend on the allocated data but also on code size.
%For energy measurements, depicted in
%Figure~\ref{fig:energy}, we compiled an application
%that first inserts 290 tuples from the BTCSAMPLE dataset, then executes 100
%find operations on the inserted tuples.
%During that process, an independent sensor node was used to monitor voltage
%and current values in short intervals (about 30ms).

\subsection{Execution Times and Energy Consumption}
\label{sec:speed}
\label{sec:energy}

We evaluated the \TS{} in terms of execution time and energy consumption of its
basic operations \texttt{insert}, \texttt{query} and \texttt{erase}.  To our
knowledge, the \TS{} is the first multi-platform embedded database optimized for
storing RDF.  As it still can be used for general tuples and thus might be
considered a general purpose embedded database, a comparison to existing
embedded databases seems natural.  We consider Antelope, a flash-based
flexible general-purpose relational database for the Contiki operating system
as well as the local, RAM-based storage of the TeenyLIME, a distributed tuple
space for TinyOS.  As all systems include the TMote Sky as possible target
platform, we use it for comparison.  The experiments were conducted on the
\textit{w.iLab.t
Testbed}\footnote{\url{http://www.crew-project.eu/portal/wilabdoc}} provided
by the CREW project\footnote{\url{http://www.crew-project.eu}}, located in the
iMinds research center\footnote{\url{http://www.iminds.be/en}}. The testbed is
equipped with 193 TMote Sky Sensor motes with 10kB RAM and 8MHz MSP430
processors. The motes are connected to a programmable system that can be used
for energy measurements and the simulation of sensor value inputs to the node.
As input data we considered the NODE dataset described in
Section~\ref{sec:datasets}.
% an RDF document as generated by LD4Sensors\todo{ref}
%for SPITFIRE describing sensor data and metadata in 73 triples.
As Antelope and TeenyLIME do not feature storage of variably-sized strings, we
configured them to account for strings of a maximum length of 120 characters,
which, as Figure~\ref{fig:lengths} illustrates, is sufficient for most
(but not all) RDF
elements in our datasets and all elements in NODE. This gives a 
bit of advantage to these implementations, as we allow them to reject
some valid data.
%which as suggested in Section~\ref{sec:rdf_properties} should be sufficient
%for the majority of RDF URIs.
TeenyLIME was given 6240 Bytes of RAM to use for Tuple storage, the Wiselib
TupleStore was configured with a tuple container with 76 elements and a
Chopper Dictionary with 100 entry slots, each 15 bytes, thus using a total of
2512 bytes of RAM.

\begin{figure*}[ht]
	\subfloat{
		\includegraphics[width=.33\textwidth]{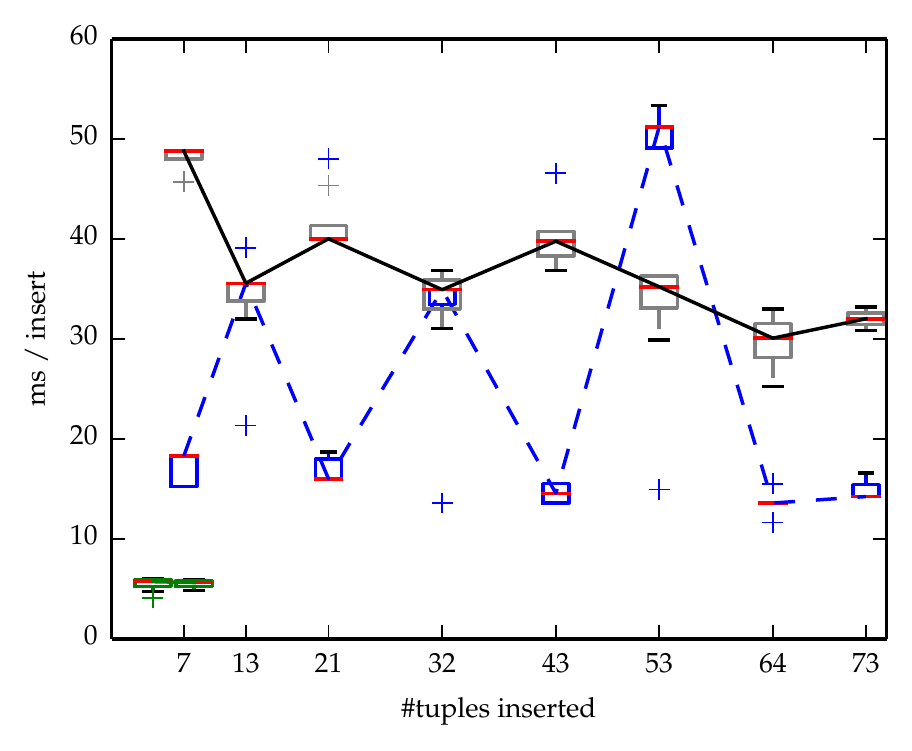}
	}
	\subfloat{
		\includegraphics[width=.33\textwidth]{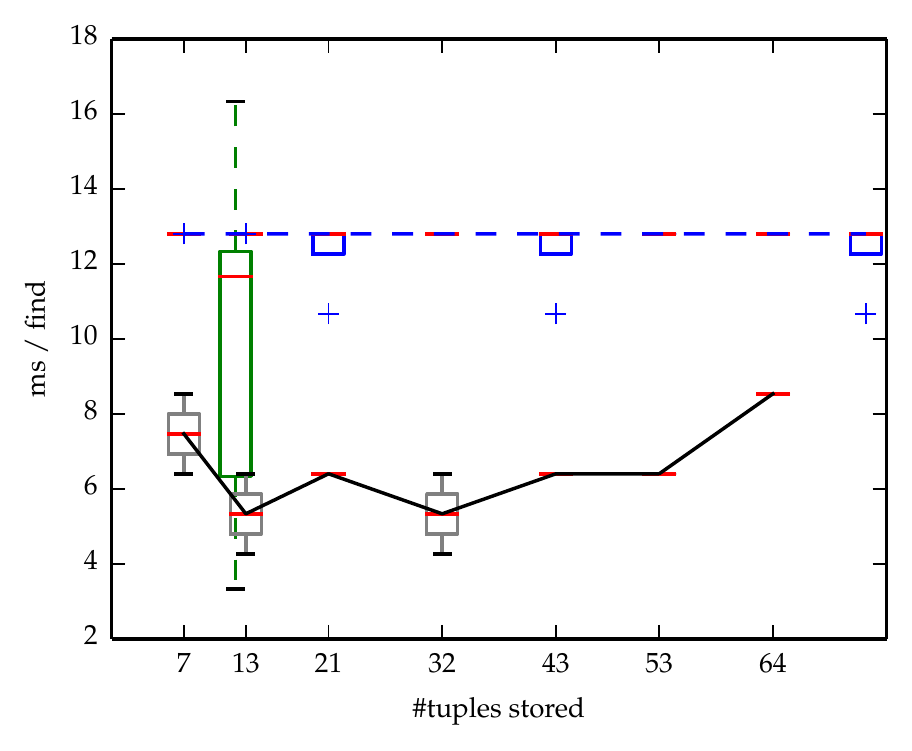}
	}
	\subfloat{
		\includegraphics[width=.33\textwidth]{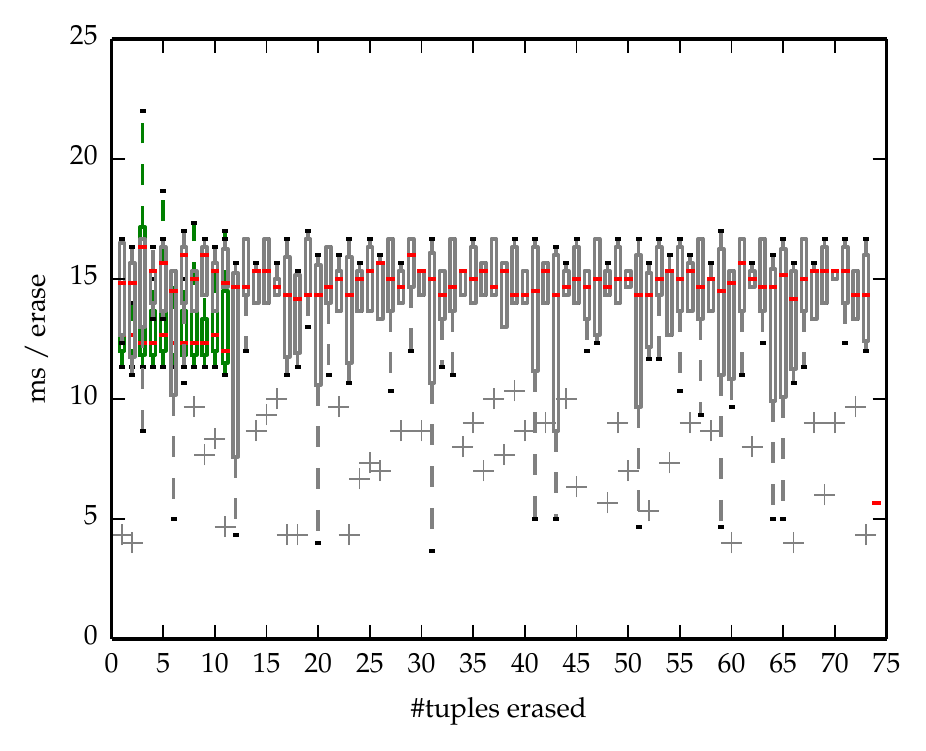}
	}
	\caption{Execution times for tuple operations for Antelope (on flash
storage, blue-dashed), TeenyLIME (in RAM, green) and the Wiselib
TupleStore (RAM, black, straight line).
Left, Center: \texttt{insert} and \texttt{query} with different number of tuples
present in the store. Right: \texttt{erase} dependant on the number of tuples
already erased (0 equals 73 tuples present in \TS{} and 12 in TeenyLIME).
}
	\label{fig:times}
\end{figure*}

\begin{figure*}[ht]
	%\begin{subfigure}[t]{.33\textwidth}
	\subfloat{
		\includegraphics[width=.33\textwidth]{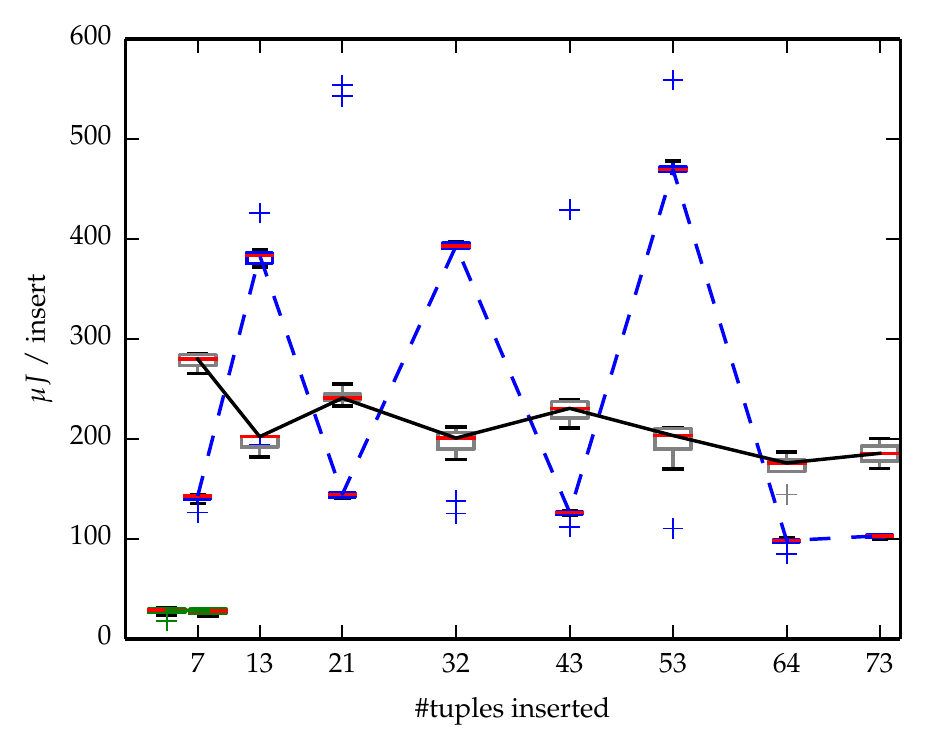}
		%\caption{Energy consumption during inserting.}
		%\label{fig:energy_insert}
	}
	%\end{subfigure}
	%\begin{subfigure}[t]{.33\textwidth}
	\subfloat{
		\includegraphics[width=.33\textwidth]{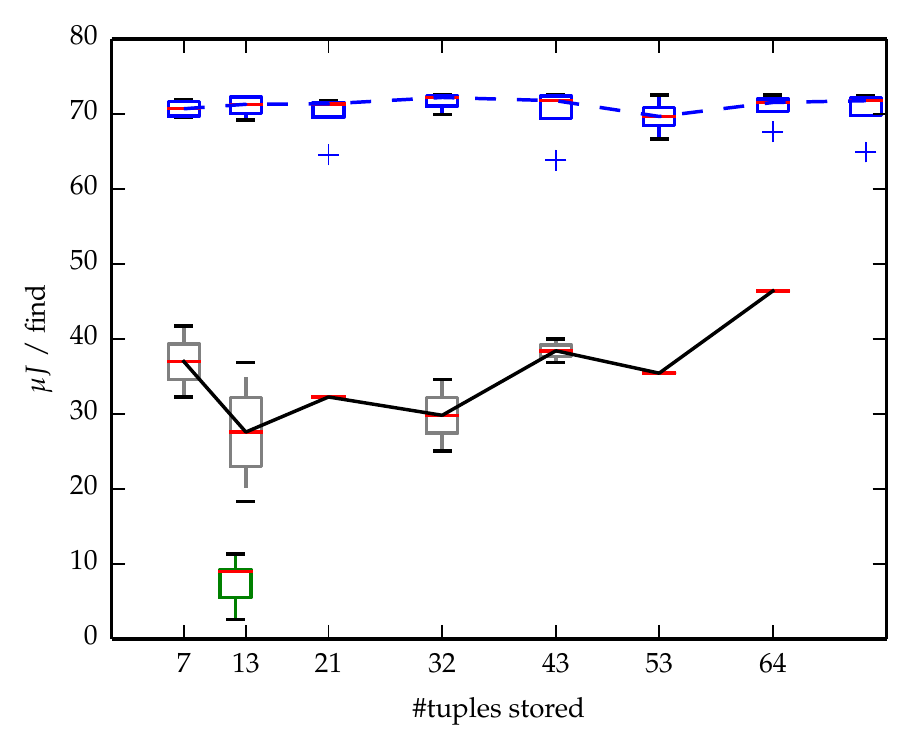}
		%\caption{Energy consumption during finding tuples using a wildcard match.}
		%\label{fig:energy_find}
	}
	%\end{subfigure}
	%\begin{subfigure}[t]{.33\textwidth}
	\subfloat{
		\includegraphics[width=.33\textwidth]{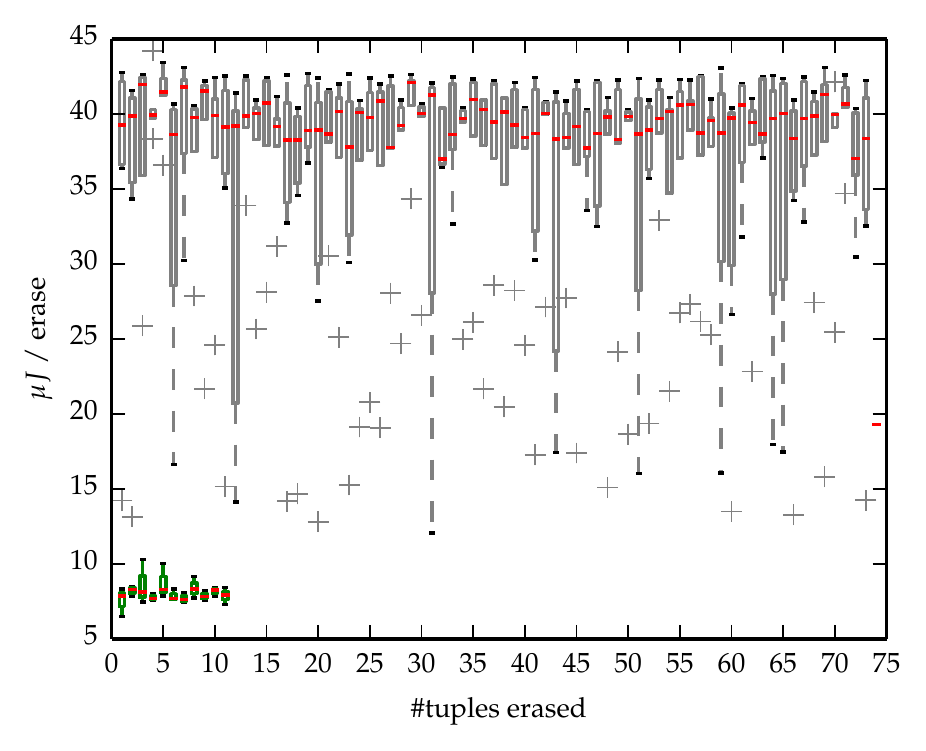}
		%\caption{Energy consumption during deletion of tuples using a wildcard match.}
		%\label{fig:energy_erase}
	}
	%\end{subfigure}
	%\caption{Energy consumption for insert- and erase operations for Antelope
%(using the builtin flash memory), TeenyLIME (in RAM) and the Wiselib
%TupleStore (RAM).}
	\caption{Energy consumption for tuple operations for Antelope (on flash
storage, blue-dashed), TeenyLIME (in RAM, green) and the Wiselib
TupleStore (RAM, black, straight line).
Left, Center: \texttt{insert} and \texttt{query} with different number of tuples
present in the store. Right: \texttt{erase} dependant on the number of tuples
already erased (0 equals 73 tuples present in \TS{} and 12 in TeenyLIME).
}
	\label{fig:energies}
\end{figure*}

In order to compensate for call overhead, triples were inserted in groups such
that their unencoded size did just not exceed 1kB. This approach had to be
slightly adapted for the TeenyLIME test to account for a smaller number of
tuples that TeenyLIME can manage.
%as we were not able to configure
%TeenyLIME to a capacity of more than 12 tuples.
With the tuples inserted we issued \texttt{query} and \texttt{erase} commands
both with random, findable tuple patterns containing one wildcard at a random position as input (e.g.
{\small\texttt{(<http://$\dots$> <http://$\dots$> *)}}).
The observed distributions of energy consumption of the
different tuple store operations are depicted in
Figure~\ref{fig:energies} and that of the execution times in
Figure~\ref{fig:times}.

%During our experiments, we observed that TeenyLIME had a significantly higher
%execution time (a median of about 12ms over 80 experiments) for finding
%tuples (that is, executing a \texttt{rd()} operation) than Antelope or the
%Wiselib Tuplestore, it is thus not shown in the figures.
%%As the implementation however seems to use the efficient \texttt{memcmp()}
%%comparison, we
%We assume that TeenyLIME -- with its focus on exchanging tuples in a
%distributed system -- triggers additional functionality here, that we were
%not able to disable.  We thus have to consider this particular result as not
%comparable to those of the other databases.
Antelope does not provide a built-in string comparison routine, thus queries
for Antelope were substituted with a simple routine that iterates over all
tuples and compares them for equality using \texttt{strcmp()}, taking
wildcards into account.  As it is not possible to select tuples by string
value equality in Antelope, we could not evaluate the erasure of tuples in
that database in a meaningful way. We note that it is possible to substitute
this missing functionality with selecting all non-matching tuples, insert them
into a new relation and then free the old relation, which is similar to what
Antelope does internally on deletion of (non-string) tuples.

The TMote Sky platform features 1024kB of external flash, composed of 16
segments a 64kB which is used in combination with some of the available RAM by
Antelope to store tuples.  The ``zig-zag'' shape of the Antelope insertion
execution time and energy consumptions can be thus be explained by caching and
flash I/O semantics (e.g. tuples being cached in RAM and only written to
external memory when there is no space left in RAM).  Furthermore, we observe
that only two sets of tuples (4 tuples each) could be inserted for the
TeenyLIME configuration, as the third group of inserts already exhausted the
available memory and was only partially inserted.  It might be possible to
tweak TeenyLIME to a more efficient use of the available memory (e.g. by
finding a good value for the memory slab size). Due to the (uncompressed) way
TeenyLIME handles string data however, it could not possibly fit more than
$\lfloor 6240 / (3 * 120) \rfloor = 17$ tuples in RAM (neglecting all
potential overhead induced by data structures and metadata).

Figures~\ref{fig:times} and~\ref{fig:energies} give an overview over execution
times and energy consumptions of the three \TS{} operations, respectively.
Due to its higher complexity and greater storage space (in terms of tuple
count), insertion in the \TS{} costs notably more time and energy than in TeenyLIME and is
comparable to Antelope (which however, operates on external flash).
In terms of lookup, TeenyLIME is notably faster, but \TS{} seems to provide the
most energy-efficient lookups.
We believe this efficiency to be related
directly with the dictionary approach: During a query, each element string
needs to be looked up only once, after it is located, the \TS{} seeks to locate
a matching tuple of dictionary keys.
In contrast in TeenyLIME, the elements of the query tuple have to be compared
to elements of all other tuples until a matching tuple is found, resulting in
higher energy consumption, the analogue holds for erasure of tuples by
template, which boils down to executing a query followed by a relatively
inexpensive delete operation.

From the experiments we conclude that there is an overhead to be paid
for using the \TS{}, however it is reasonably small. Our database for
arbitrary string-based RDF data uses about as much energy and runtime
as much simpler tuple spaces with fixed record size, but opens the
door to much richer applications.

%%% Local Variables: 
%%% mode: latex
%%% TeX-master: "article"
%%% End: 

\section{Conclusion}
\label{sec:conclusion}
By enabling embedded devices to describe their state and their observations in
the universal RDF format, these devices can be integrated into the Semantic
Web.
This allows not only for easy machine-to-machine communication but also (in
conjunction with other components) for posing queries that combine information
from the sensor network and documents in the web.
To show that this idea is feasible, we introduced the \TS.
The \TS\ provides a potent mechanism for managing arbitrary RDF data on resource-constrained
embedded devices. Due to its portability and modularity it can be set up to utilize flash
memory or RAM and adopt to the capabilities and resource constraints of almost any
given device.
In experiments comparing it against TeenyLIME and Antelope, we have
shown that energy and runtime cost is reasonable, especially
considering that our implementation can work with arbitrary data, and
runs without code porting on many different platforms.
% We have shown that our implementation is RAM-based implementation can store
% RDF much more efficiently than TeenyLIME and query it more efficiently than
% Antelope.

%\section*{Acknowledgments}

%We would like to thank Matteo Ceriotti for helpful comments and
%Marcus Brandenburger and Ulf Kulau for design and implementation of
%energy measurement hardware for first experiments.
%The authors would also like to thank the people maintaining the the w.iLab.t
%testbed for their unwavering technical support.
%This work has been partially supported by the European Union (ICT-2009-258885, SPITFIRE).

%\bibliographystyle{IEEEtran}
\bibliographystyle{abbrv}
\bibliography{references}

% that's all folks
\end{document}